\documentclass{PoS}
\usepackage{graphicx}                            

\def\DU  {\mathop{{\cal D}\hbox{U}}}

\def\Dxi {\mathop{{\cal D}\xi}}
\def\dd  {\mbox{d}}
\def\Tr  {\rm Tr}
\def\MeV  {\rm MeV}

\def\Re {\mathop{\hbox{Re}}}
\def\beq {\begin{equation}}
\def\eeq {\end{equation}}
\newcommand\comment[1]{}

\title{Finite density simulations using a determinant estimator}
\ShortTitle{Finite density simulations using a determinant estimator}

\author{\speaker{Andrei Alexandru}, Anyi Li, Keh-Fei Liu\\
        Department of Physics and Astronomy, University of Kentucky, Lexington KY 40506, USA\\
        E-mail: \email{alexan@pa.uky.edu}, \email{anyili@pa.uky.edu}, \email{liu@pa.uky.edu}}

\abstract{
Previous investigations have shown that the canonical approach 
to simulating QCD at finite density is promising. The algorithm 
we used in our earlier work employs an exact calculation of the 
fermionic determinant which limits the size of the lattices 
we can simulate. Interesting questions can only be answered 
if we simulate at larger volume. In this paper we explore an 
algorithm, Hybrid Noisy Monte Carlo, that employs a determinant 
estimator rather than an exact calculation. We first present 
the technical aspects of the estimator, check that the algorithm 
is correct by comparing it with our previous study, and then 
discuss its merits. We will also discuss the challenges faced 
when simulating larger lattice volumes. 
}

\FullConference{The XXV International Symposium on Lattice Field Theory\\
         July 30-4 August 2007\\
         Regensburg, Germany}

\begin{document}

\section{Introduction}
Simulating QCD at non-zero baryon density remains one of the challenges of
lattice QCD. Direct Monte Carlo simulations are hindered by the fact that
the fermionic determinant is complex when the chemical potential is different
from zero. The common solution of separating out a complex phase from the
integrand and introducing it in the observable fails due to the sign problem
and overlap problem. A solution to the overlap problem was proposed in
\cite{kfl05}; the method proposed starts from the canonical partition function
rather than the usual approach based on the grand canonical one.
For two degenerate flavors of quarks the canonical partition function is:
\beq
Z_C(V, T, k) = \int \DU e^{-S_G(U)} {\det}_k M(U)^2,
\eeq
where
\beq
{\det}_k M(U)^2 \equiv \frac{1}{2\pi} \int_0^{2\pi} \dd\phi e^{-i k \phi} \det M(U, \mu=i\phi T)^2.
\eeq
With the above choice the total net quark number is fixed, i.e. $n_u + n_d = k$.
Simulating the above action involves computing the determinant for every 
phase $\phi$; this is not feasible. 
We replace the continuous Fourier transform, ${\det}_k M^2$ with a discrete one,
${\det}'_k M^2$.
The error due
to this approximation is discussed in \cite{aa05} and further analyzed in \cite{al07}.
In the canonical approach we still have to integrate over a complex integrand; the
fermion contribution ${\det}_k M(U)^2$ is complex when $k\not=0$. In our study
\cite{aa05} we used a Monte Carlo method to generate an ensemble based on the weight
\beq
W(U) \propto e^{-S_G(U)} \left| \Re {\det}'_k M(U)^2 \right|,
\eeq
and the removed phase
is reintroduced in the observable measurement. To generate an ensemble with this
weight we use Metropolis method where the accept/reject step is based on the 
ratio of determinants. For more details the reader is referred to the original 
paper \cite{aa05}. 

Calculating the determinant of the fermionic matrix exactly is time consuming; 
it is only feasible for small lattices. To move to larger lattices we will use
a method that involves an estimator for the determinant. This method was
proposed in \cite{kfl05}; in this work we will investigate it numerically.

\section{The algorithm}
In this section we will present the algorithm used in our simulations.
We start by rewriting the partition function
\begin{eqnarray}
Z_C(V,T,k) &=& \int \DU e^{-S_G(U)} {\det}'_k M(U)^2 \\ \nonumber
 &=& \int \DU\Dxi e^{-S_G(U)} \det M(U)^2 f_k(U,\xi),
\end{eqnarray}
where
\beq
\int \Dxi g(U, \phi,\xi) = \frac{\det M(U_\phi)^2}{\det M(U)^2},
\label{eq-estimator}
\eeq
and
\beq
f_k(U,\xi)=\frac{1}{N}\sum_{\phi_i} e^{-i k\phi_i}g(U, \phi_i, \xi).
\eeq
We introduce the estimator $f_k(U,\xi)$ by rewriting the integral in terms
of a new, auxiliary field $\xi$. 
This auxiliary field plays the role of
a stochastic variable. The other important feature is that we separate out
$\det M(U)^2$ and we estimate the ratio ${\det}'_k M(U)^2/\det M(U)^2$.
There are two reasons for this separation: it improves the acceptance rate and
also the ratio estimator is more precise than the estimator for the determinant
alone.

Since the determinant projection is complex its estimator will also be complex.
In order to simulate it, we need to separate a positive part out and fold the 
remaining phase in the observables. We will then generate ensembles based on the
weight:
\beq
W(U, \xi) \propto e^{-S_G(U)} \det M(U)^2 \left| f_k (U, \xi)\right |.
\label{eq-weight}
\eeq

The algorithm proceeds in two steps: we first update the gauge fields 
$U\rightarrow U'$ using an HMC update that satisfies detailed balance
with respect to $e^{-S_G(U)} \det M(U)^2$ and accept this change with the
probability baseon on the ratio $\left| \frac{f_k(U', \xi)}{f_k(U,\xi)} \right|$.
The reason for separating out $\det M(U)^2$ is that
the new gauge field proposal takes into account the fermionic contribution. As
showed in \cite{aa05} this improves the acceptance rate of this step dramatically.

The second step in the updating process is to refresh the auxiliary variables $\xi$.
To implement the Noisy Monte Carlo algorithm~\cite{lls00}, another
accept/reject step is used where a new proposal $\xi\rightarrow\xi'$ is accepted 
with probability based on the ratio $ \left| \frac{f_k(U, \xi')}{f_k(U,\xi)} \right|$.
It is easy to check that this algorithm satisfies the detailed balance with respect to
the weight $W(U,\xi)$ in Eq. (\ref{eq-weight}).

\section{The estimator}

The most interesting technical aspect is setting up the estimator. In this section
we describe how to set up $g(U,\phi, \xi)$, an unbiased estimator for the determinant
ratio~\cite{ct98}. We start by rewriting Eq. (\ref{eq-estimator}):
\beq
\int \DU\Dxi g(U, \phi, \xi)= e^{2\Tr (\ln M(U_\phi) - \ln M(U))}.
\label{eq-31}
\eeq
The construction of the estimator occurs in steps. First approximate
$\ln M$ using a Pade approximation:
\beq
\ln M \approx b_0 I + \sum_{i=1}^{K} \frac{b_i}{c_i + M}
\label{eq-lnexpansion}
\eeq
For our simulations, we used a Pade approximation of order $K=30$.

\begin{figure}[th]
  \centering
  \includegraphics[height = 5cm]{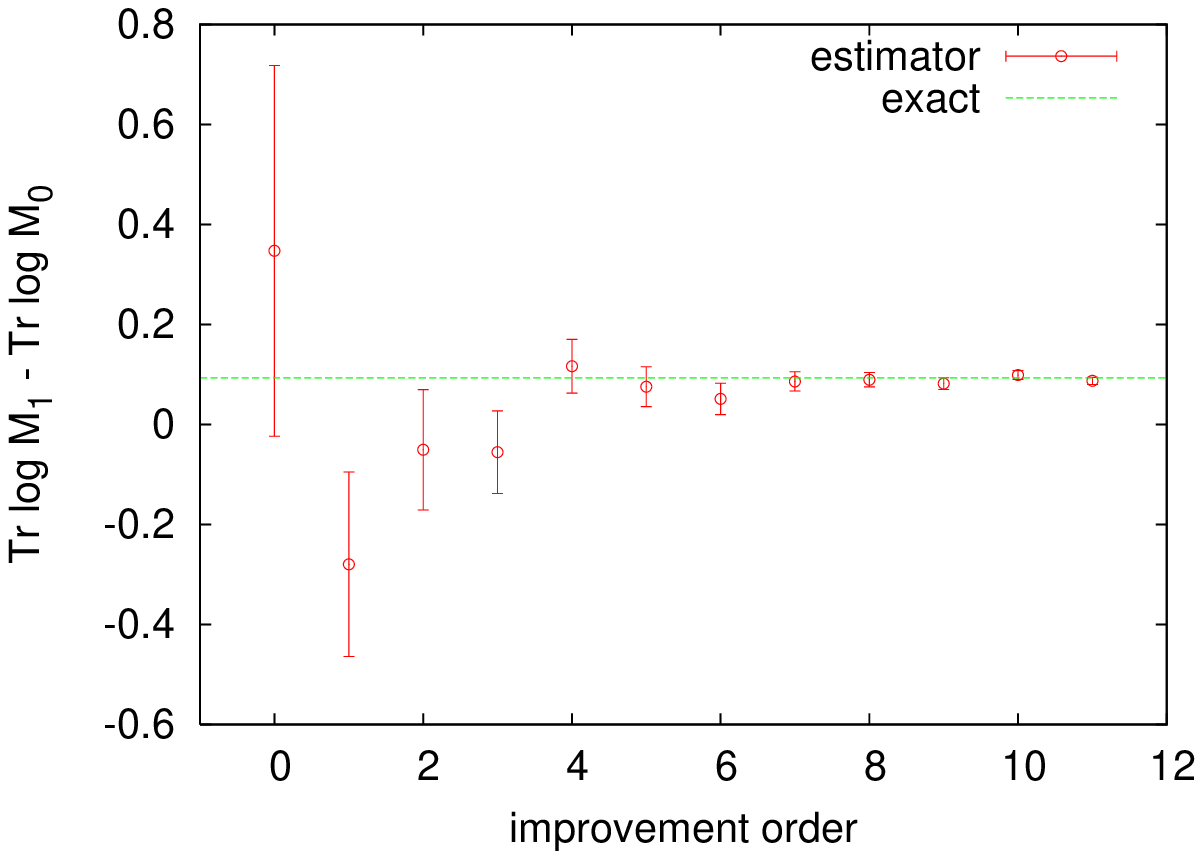}
  \includegraphics[height = 5cm]{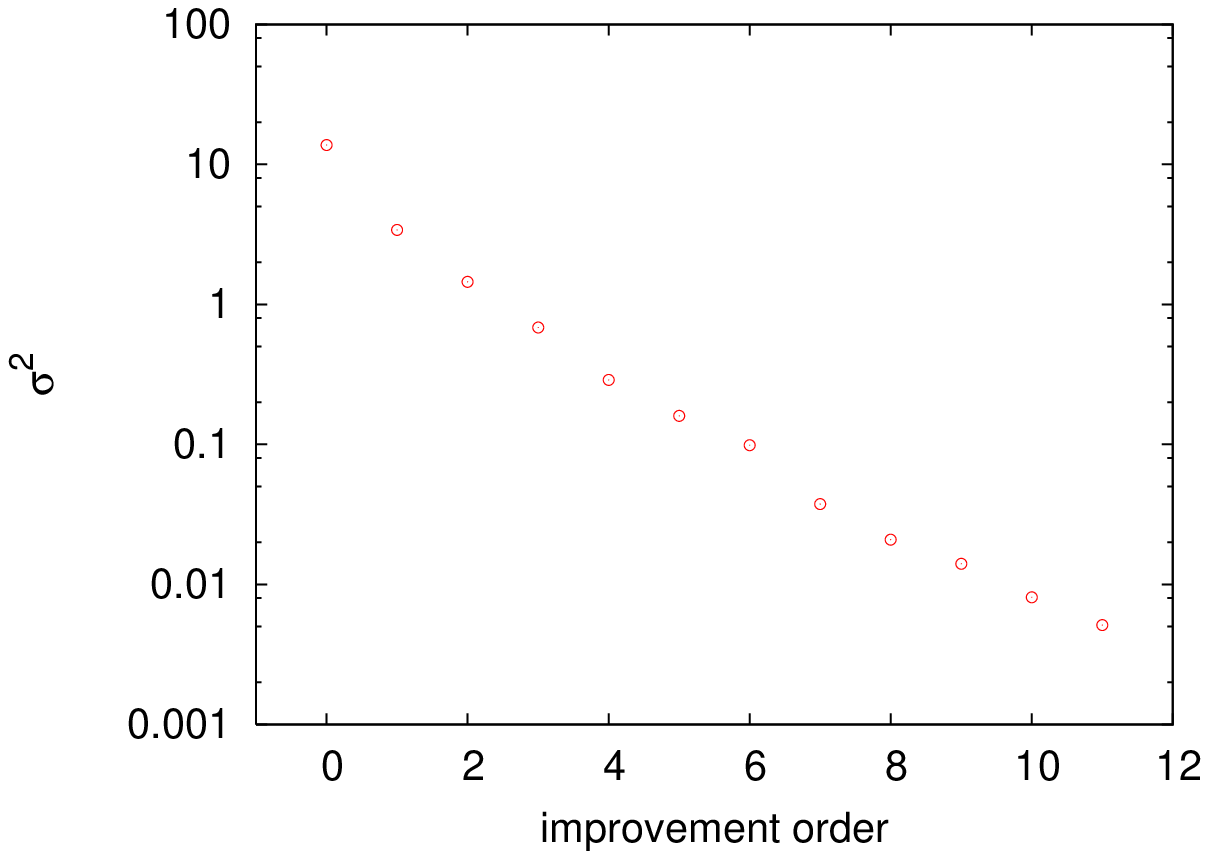}
  \caption{Improvement of the trace estimator. Left panel: average and error
of 1000 different noises with different level of improvement; the line corresponds
to the exact value. Right panel: the variance as a function of the improvement order. }
  \label{fig-improvement}
\end{figure}

We then set up an estimator for the exponent of the right hand side of
Eq. (\ref{eq-31}). For the trace, we use an unbiased estimator based on $Z(4)$
noise. We generate random vectors $\eta$ and use 
\beq
h(\eta) = 2\eta^\dagger [\ln M(U_\phi) - \ln M(U)]\eta,
\eeq
as an estimator for $2\Tr (\ln M_\phi -\ln M)$. 
It is easy to show that when the elements
of $\eta$ are picked with equal probability from $Z(4)=\{1, -1, i, -i\}$, the
estimator $h(\eta)$ is unbiased. Unfortunately, without improvement this estimator
has a large variance. In \cite{ct98} it is shown that the variance is
proportional with the magnitude of the off-diagonal elements of the matrix.
The suggested strategy is to subtract from the matrix of interest matrices 
that are traceless, so that the estimator remains unbiased, and that
emulate the off-diagonal structure of the original matrix, so that the variance 
is reduced. 
We carry out this program using the set of matrices $D^k$, where $D$ is
the Wilson hopping matrix. The order of the improvement is given by the
largest $k$ in the set. 
The variance is
reduced exponentially with the order of improvement as can be seen from
Fig. \ref{fig-improvement}. The only problem is that $\Tr D^k$ is non-zero
when $k$ is even and larger than 2; for these orders we use the traceless
matrix $D^k - \frac{1}{N} \Tr D^k$, where $N$ is the dimension of the matrix.
The computational burden is then to compute $\Tr D^k$; to evaluate this we
have to compute all closed loops of $k$ steps. For $k=4$, we have $6$ such
loops at every point corresponding to different plaquette orientations and,
on a $4^4$ lattice, $4$ Polyakov loops wrapping around the lattice in different
directions. As we increase $k$ the number of such loops increases quickly,
$112$ for $k=6$, $2884$ for $k=8$ and $84360$ for $k=10$. For the purpose of
testing, we carried out the calculation up to order $k=11$, but this is very
expensive. In our simulations we used only improvement up to order 9.

Once the estimator for the exponent is set up, we have to design an unbiased 
estimator for the exponential function. Using the exponent estimator $h$,
we follow \cite{gb85} and write

\beq
g[h](\eta_1, \eta_2, ...) = 1 + h(\eta_1) + 
H(\theta_1-\frac{1}{2}) h(\eta_1)h(\eta_2) +
H(\theta_1-\frac{1}{2})
H(\theta_2-\frac{2}{3}) h(\eta_1)h(\eta_2)h(\eta_3) + ...,
\label{eq-gdef}
\eeq
where $H(x)$ is the step function which is $0$ for $x<0$ and equals $1$ otherwise.
The numbers $\theta_i$ are random variables uniformly distributed over the
interval $[0, 1]$. The series looks infinite, but as soon as one of the step
functions is zero we can stop evaluating the following terms. In general 
this happens quite fast and the average number of exponent estimators 
$h(\eta_i)$ that needs to be evaluated is $e-1\approx 1.72$. It is easy to
show that this estimator averages to the desired value.

\begin{figure}[th]
  \centering
  \includegraphics[height = 4.75cm]{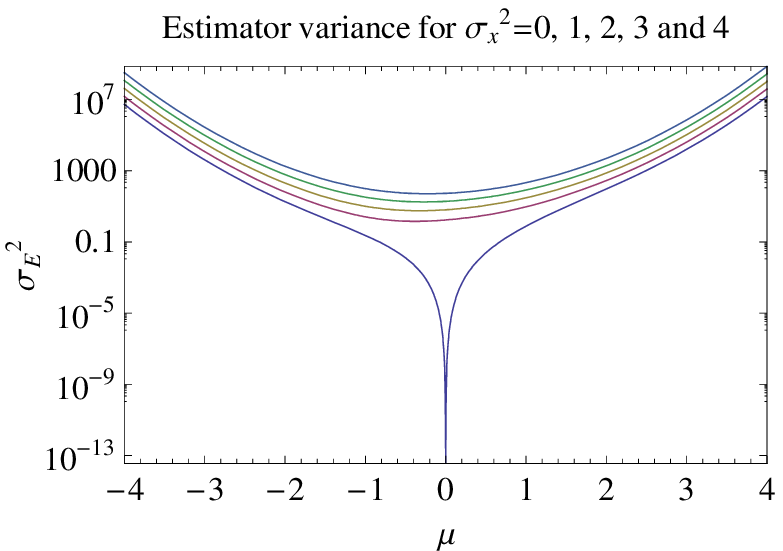}
  \includegraphics[height = 4.8cm]{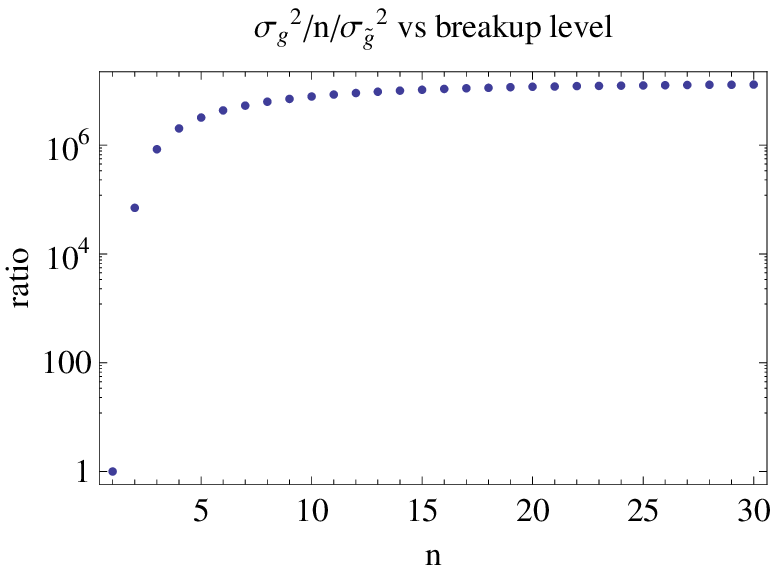}
  \caption{Exponential estimator variance. 
Left panel: $\mu$ denotes the average value of the
$h$ estimator and $\sigma_x^2$ its variance; the variance of the $g$ estimator
is denoted by $\sigma_E^2$. Right panel: we plot the increase in variance
when using naive averaging of $n$ estimates of $g$ compared to the variance
of $\tilde{g}$.}
  \label{fig-varest}
\end{figure}

We have now an unbiased estimator for $\frac{\det M_\phi^2}{\det M^2}$. The
only problem is that it has a very large variance. To understand how to 
address this problem, we plot in Fig. \ref{fig-varest} the variance of the
estimator $g$ as a function of the mean value and variance of the estimator $h$.
You can see that the variance of the estimator grows very quickly when the
mean value or the variance of $h$ are larger than $1$. Unfortunately, we cannot
control how big these quantities are; they are dictated by how large the 
lattice is and the ensemble temperature. 

To address this problem, we use
an estimator breakup: imagine that instead of $h(\eta)$ in Eq. (\ref{eq-gdef})
we use $h(\eta)/n$. Then the $g$ estimator would average to the $n^{\rm th}$
root of the determinant ratio. If we take $n$ independent estimates of this
new $g$ and multiply them together we get a new estimator for the ratio $\tilde{g}$.
This estimator has a smaller variance than $g$, but it is $n$
times more expensive to evaluate. To confirm that this trick is useful, 
we compare the variance decrease with the one that we would get by just 
evaluating the average of $n$ independent estimates of $g$. This average
costs the same as $\tilde{g}$ and its variance is $n$ times smaller
than the variance of $g$. In Fig. \ref{fig-varest} we plot this ratio
(i.e. $\frac{\sigma_g^2/n}{\sigma_{\tilde{g}}^2}$)
for a case when $\left< h\right> = 5$ and  $\left<h^2\right> = 5.1$ 
(small variance for the exponent).
We see from the figure that if we use the naive averaging, the variance would
be increased by about a million times when the breakup level is greater than
$5$. 

The first lesson is that the breakup estimator has a much better variance
than the native averaging. The second important thing to notice is that 
when the breakup level is greater than the average value of $h$ the
payoff levels off; this means that further reductions in variance come
at the same rate as statistical reduction. We derive then a rule of the
thumb to help us in setting up the breakup level: {\em the breakup level, $n$,
should be set so that most of the estimates of $h$ are smaller than $n$}. In 
practice this means that we have to setup $n$ larger than the average
value of $h$ plus a few standard deviations.

\section{Algorithm check}

To verify that the algorithm is correct we run a set of simulations on
$4^4$ lattices at the same parameters as the ensembles we generated in
our previous work \cite{aa05}. 
Before we present the results, we want to discuss the simulations. We had
run simulations at $\beta=5.10$, $5.15$, $5.20$, $5.25$ and $5.30$.
These simulations 
correspond to temperatures between $153\, \MeV$ and $189\, \MeV$. We
chose these simulation points since they are close to the transition
temperature and this is the area that we are interesting in scanning in
our future work. For each temperature, we ran three simulations for zero,
one and two baryon numbers. 
The hopping parameter was set to $\kappa = 0.158$ which
corresponds to a pion mass of approximatively $1\, {\rm GeV}$. We tuned the 
HMC trajectories length so that the acceptance rate of the gauge update
is about 50\%.

The only new parameter that we had to set up is the breakup level. As
discussed in the previous section, the breakup level needs to be setup
such that most of the estimates are smaller than $n$. To simplify things we
took the ensembles we generated in our previous study and, for each
configuration, we evaluated the determinant for all the phases we 
used in that study. For each configuration, we computed 
$\max_\phi \Tr\ln M_\phi - \min_\phi \Tr\ln M_\phi$ and we found it
to be bound by $15$. To be on the safe side, we set the breakup level
to $n=20$.

\begin{figure}[th]
  \centering
  \includegraphics[height = 4.8cm]{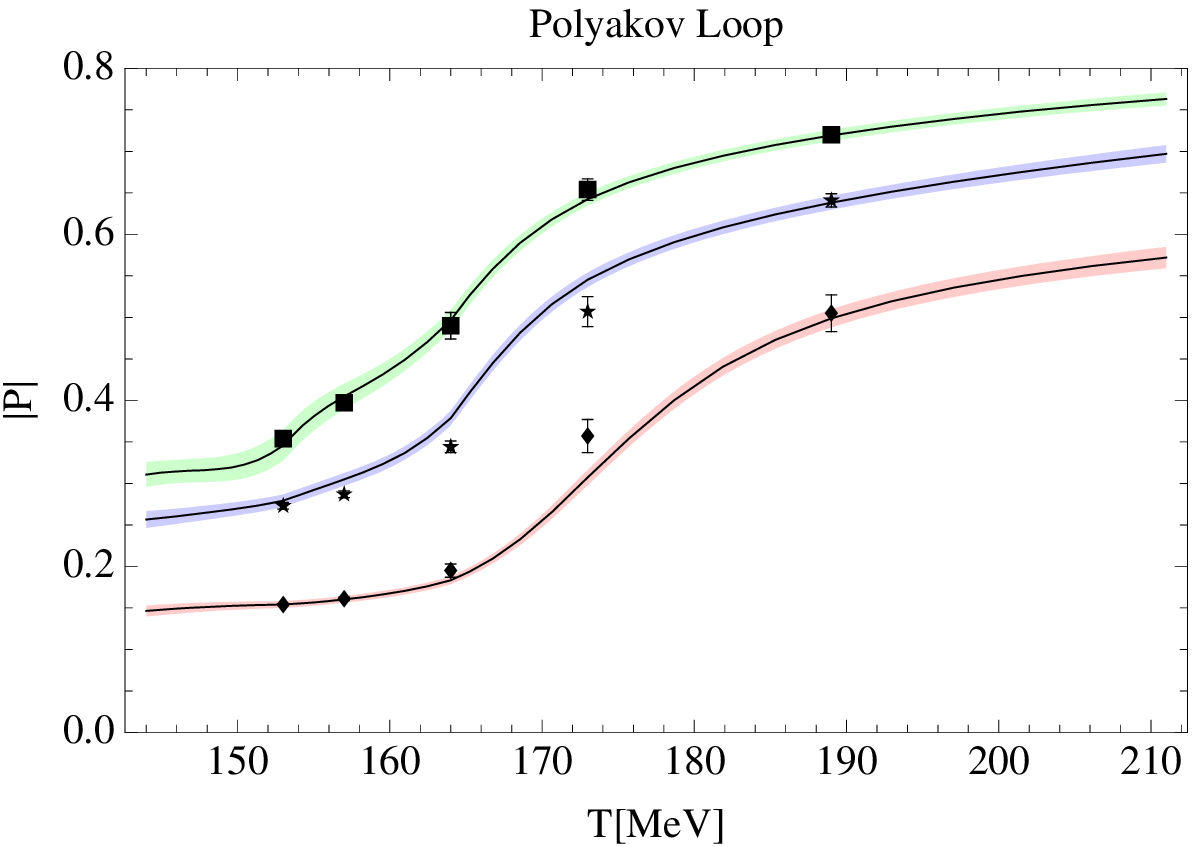}
  \includegraphics[height = 4.8cm]{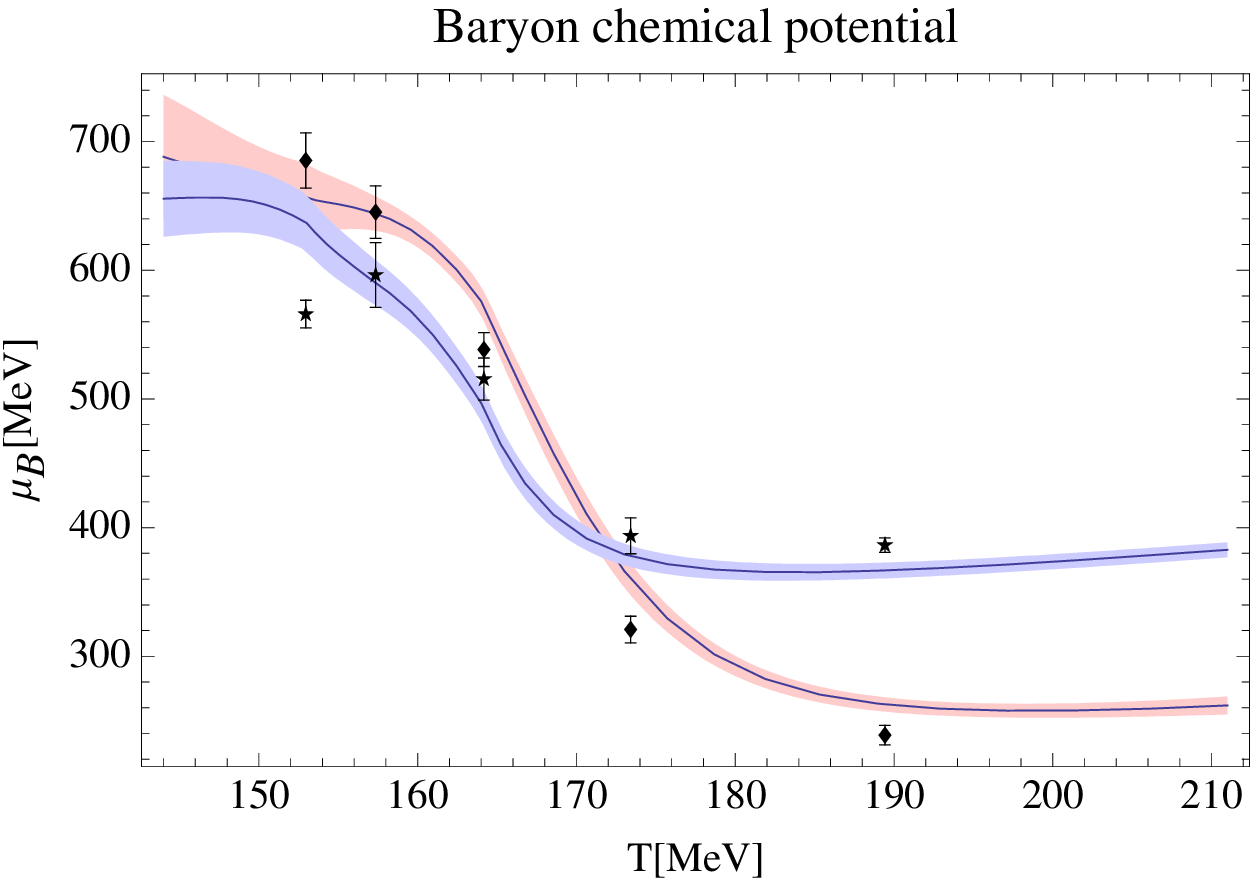}
  \caption{Checking the algorithm. 
Left panel: Absolute value of the Polyakov loop; the solid lines are
the result of our previous work and the symbols are the result of our
current simulation. From bottom up the curves correspond to 0, 1 and
2 baryon simulations.
Right panel: Chemical potential -- the lines
are our previous results and the symbols are the results of our current
simulations. The red curve represents the difference in free energy 
between the 1 and 0 baryon simulations and the blue curve corresponds
to the difference between 2 and 1.} 
  \label{fig-compare}
\end{figure}

To compare the results of the current simulations with the ones
employing an exact calculation of the determinant, we chose to 
measure the absolute value of the Polyakov loop and the baryon
chemical potential. 
In Fig. \ref{fig-compare} we compare the results of our
runs with the results we got from our previous study; the results
agree quite well and this leads us to conclude that the method
and implementation are sound. 

\section{Volume dependence}

Our first simulations on $4^4$ lattices confirm the correctness of
the algorithm. However, to produce interesting results we need to 
move to larger volumes. The main drive behind the idea of using an
estimator in our simulation is that the exact calculations of the
determinant scale quite poorly with the size of the lattice -- as
we increase the lattice volume $V$ the cost of the calculation 
increases with $V^3$ whereas the estimator calculation was expected to 
scale like $V$ (the cost of computing $(M+b_i)^{-1}\eta$ in Eq.
(\ref{eq-lnexpansion}) needed to approximate $\ln M\eta$). 

\begin{figure}[th]
  \centering
  \includegraphics[height=4.5cm]{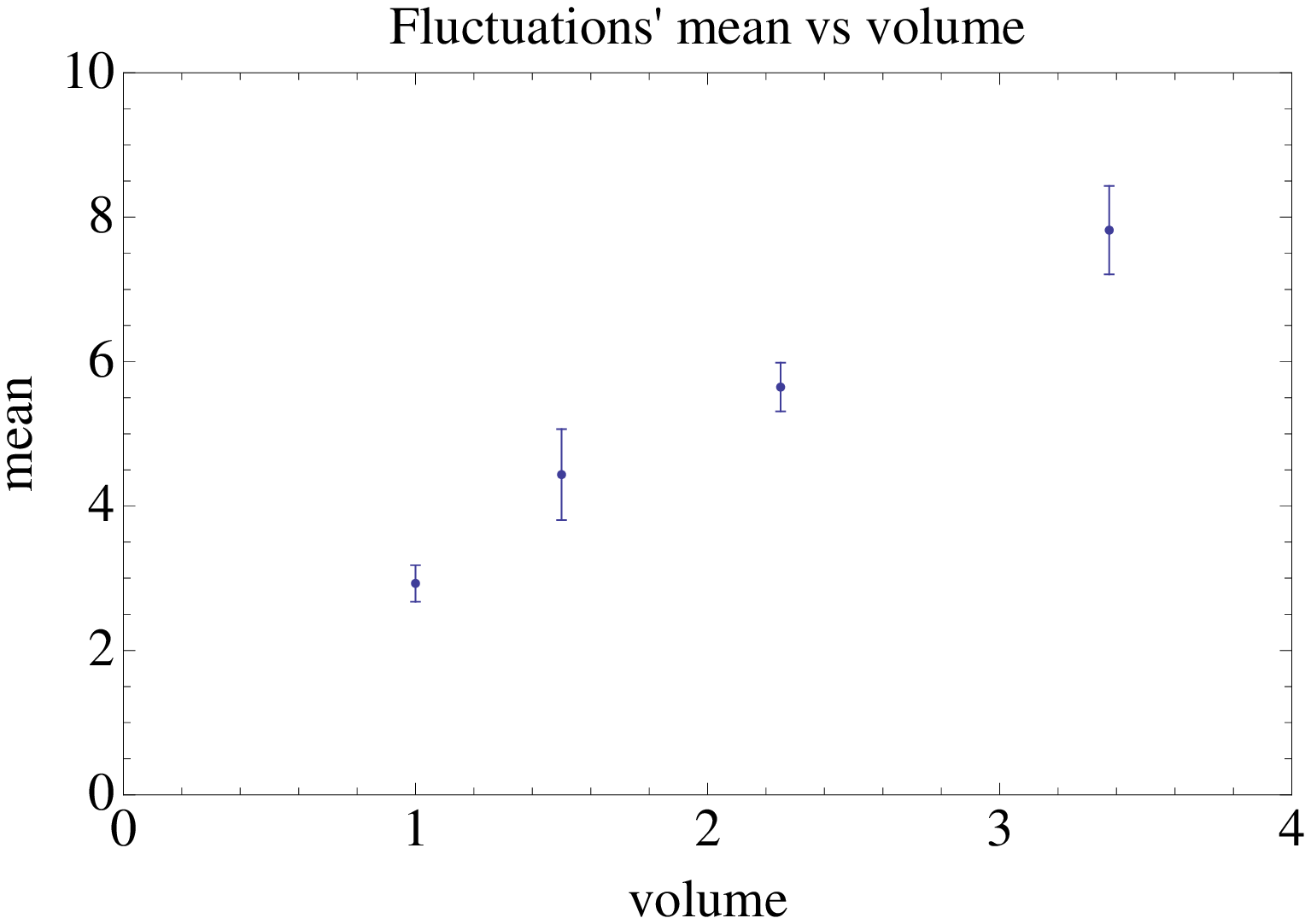}
  \includegraphics[height=4.5cm]{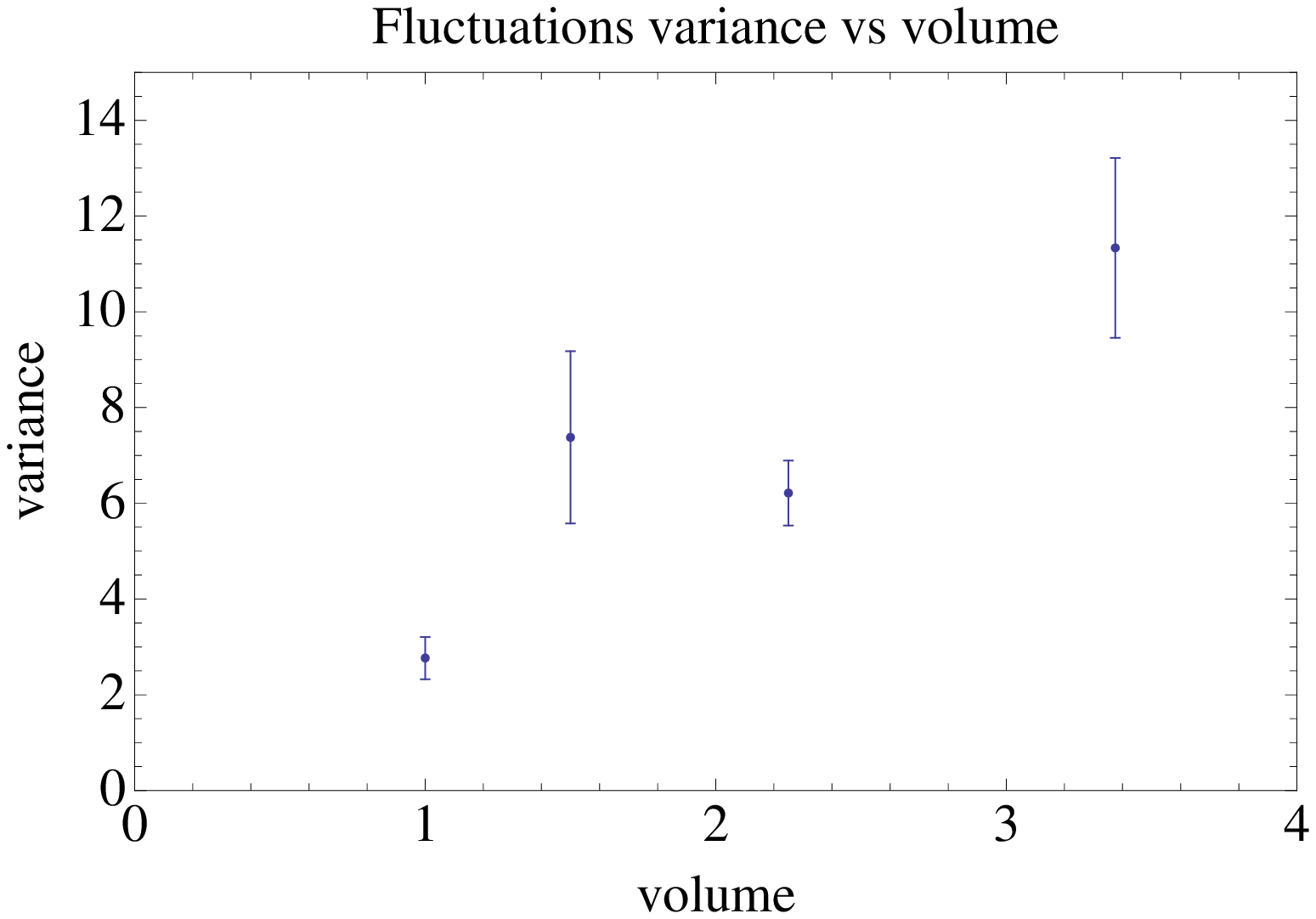}
  \caption{Volume dependence of the fluctuations.
Left panel: the mean value of the fluctuations as a function of volume.
  Right panel: the variance of the fluctuations as a function of volume.
\label{fig-voldep}}
\end{figure}

It turns out that the estimator is actually more expensive due
to the determinant breakup: the cost of the calculation increases 
linearly with the level of breakup and, as we discussed in a previous 
section, the level of breakup needs to be adjusted as the fluctuations 
of $\Tr\ln M_\phi$ become larger. To determine how these fluctuations 
scale, we ran simulations on $4^4$, $6\times 4^3$, $6^2\times 4^2$ and 
$6^3\times 4$ lattices. In Fig. \ref{fig-voldep} we show how the average 
and the variance of the fluctuations of the exponent $h$ vary with volume.
It is easy to
see that they increase linearly with the volume. We conclude then
that the cost of computing the estimator scales with $V^2$. 

\section{Conclusions}

This work is part of a larger effort aimed at simulating QCD at non-zero
baryon density. We chose to investigate an approach based on the canonical 
partition function mainly because this method should not have an overlap
problem. Previous investigations \cite{aa05, pf06} have shown it to be
very promising. 
In this paper, we analyze a method based on an estimator \cite{kfl05}. 
The main technical issue we discuss is the implementation of the estimator: 
we describe the details of our implementation and show that it
is correct by running simulations on $4^4$ lattices. The results of these 
simulations compare well with the ones we obtained using a different method 
\cite{aa05}. 

We also look at the performance of the estimator and how it 
scales with the volume. We find that the estimator is more expensive that 
originally expected. However, even on $4^4$ lattice the algorithm performs 
better than the exact method: 100 seconds for computing the estimator twice 
(once for the gauge update and once for the stochastic field update)
versus 140 seconds for the exact calculation. Furthermore, when moving
to larger lattices the gains will increase since the estimator calculation
scales as $V^2$ compared to $V^3$ for the exact calculation.

Our next goal is to run simulations on $6^3\times 4$ lattices and look for 
a phase transition signal at non-zero baryon density. This was shown to
be feasible \cite{pf06} and we plan to perform a similar analysis for
our simulations. 
In this paper we show
that the best method to attack larger volume simulations is the estimator
method and we hope to report soon on simulations at larger volumes.

\end{document}